\documentclass{skaox2006}
\usepackage{graphicx}
\begin{document}
   \title{MESMER: MeerKAT Search for Molecules in the Epoch of Reionization}

   \author{I. Heywood\inst{1}, R. P. Armstrong$^{1}$, R. Booth$^{2}$, A. J. Bunker$^{1}$, R. P. Deane$^{1}$, M. J. Jarvis$^{3,4}$, J. L. Jonas$^{5}$, M. E. Jones$^{1}$, 
   H-R. Kl\"{o}ckner$^{1,6}$, J-P. Kneib$^{7}$,
    K. K. Knudsen$^{8,9}$, F. Levrier$^{10}$, D. Obreschkow$^{1}$, D. Rigopoulou$^{1}$, S. Rawlings$^{1}$, O. M. Smirnov$^{11}$, A. C. Taylor$^{1}$, A. Verma$^{1}$,
   J. Dunlop$^{12}$, M. G. Santos$^{13}$, E. R. Stanway$^{14}$, 
          \and
           C. Willott$^{15}$
          }

   \institute{$^{1}$ Subdepartment of Astrophysics, University of Oxford, Denys-Wilkinson Building, Keble Road, Oxford, OX1 3RH, UK \\
	$^{2}$ SASKA Project, Hartebeesthoek Radio Astronomy Observatory, P.O.Box 443, Krugersdorp 1740, South Africa\\
	$^{3}$ Centre for Astrophysics Research, STRI, University of Hertfordshire, Hatfield, AL10 9AB, UK\\
	$^{4}$ University of the Western Cape, Department of Physics, 7535 Belliville, Western Cape Province, Cape Town, South Africa\\
	$^{5}$ Department of Physics \& Electronics, Rhodes University, Grahamstown, South Africa\\
	$^{6}$ Max-Planck-Institut f\"{u}r Radioastronomie, Auf dem H\"{u}gel 69, 53121 Bonn, Germany\\
	$^{7}$ Laboratoire d'Astrophysique de Marseille, CNRS- Universit\'e Aix-Marseille, 38 rue Fr\'ed\'eric Joliot-Curie, 13388 Cedex 13 Marseille, France\\
	$^{8}$ Argelander-Institut f\"{u}r Astronomie, Auf dem H\"{u}gel 71, 53121 Bonn, Germany\\
	$^{9}$ Department of Earth \& Space Sciences, Chalmers University of Technology, Onsala Space Observatory, SE-43992 Onsala, Sweden\\
	$^{10}$ Laboratoire de Radioastronomie, D\'{e}partement de Physique, \'{E}cole Normale Sup\'{e}rieure de Paris, 24 Rue Lhomond, 75231 Cedex 05 Paris, France\\
	$^{11}$ Netherlands Institute for Radio Astronomy (ASTRON), P. O. Box 2, 7990 AA Dwingeloo, The Netherlands\\ 
	$^{12}$ Institute for Astronomy, University of Edinburgh, Royal Observatory, Blackford Hill, Edinburgh EH9 3HJ, UK\\	
	$^{13}$ CENTRA, Departamento de F\'{i}sica, Instituto Superior T\'{e}cnico, 1049-001, Lisboa, Portugal \\
	$^{14}$ Department of Physics, University of Warwick, Gibbet Hill Road, Coventry, CV4 7AL, UK\\
	$^{15}$ Herzberg Institute of Astrophysics, National Research Council, 5071 West Saanich Rd, Victoria, BC V9E 2E7, Canada\\
	}

   \abstract{Observations 
of molecular gas in galaxies at all redshifts are critical
for measuring the cosmic evolution in molecular gas density 
($\Omega_{H_{2}}(z)$) and understanding the star-formation history of 
the Universe. The $^{12}$CO molecule ($\nu_{rest}$ for the $J$=1$\rightarrow$0 rotational transition = 115.27 GHz) 
is the best proxy for 
extragalactic molecular hydrogen (H$_{2}$), which
is the gas reservoir from which star formation occurs. 
The detection of CO at high-redshift has been a growing industry over the last few years with 
successful molecular gas detections out to z$\sim$6. Typically, redshifted high-$J$ transitions are 
observed using mm-wave instruments, with the most commonly targeted systems being those with high star 
formation rates such as submm galaxies, and far-infrared-bright quasars.
While the most luminous objects are naturally the most readily observed, 
observations of
the typical members of the galaxy population which exhibit modest 
star-formation rates are essential for completing the picture.
The arrival of ALMA will be revolutionary in 
terms of increasing
the detection rate and pushing the sensitivity limit down to include 
more normal star-forming galaxies, however the limited field-of-view when observing
at such high frequencies makes it difficult to use ALMA for studies of the large-scale 
structure traced out by molecular gas in galaxies. This article introduces a strategy for a 
systematic search for molecular
gas during the Epoch of Reionization (z$\sim$7 and above),
capitalizing on the fact that the $J$=1$\rightarrow$0 transition of $^{12}$CO enters 
the upper frequency bands of cm-wave instruments at the appropriate 
redshift. The
field-of-view advantage gained by observing at such frequencies, 
coupled with modern broadband correlators allows significant 
cosmological volumes
to be surveyed on reasonable timescales.
In this article we present an overview of our proposed observing 
programme which has been awarded 6,500 hours as one of the Large Survey 
Projects for MeerKAT, the forthcoming
South African Square Kilometre Array pathfinder instrument. Its large 
field of view and correlator bandwidth, and high-sensitivity
provide unprecedented survey speed for such work. An existing 
astrophysical simulation is coupled with instrumental considerations
to demonstrate the feasibility of such observations and predict 
detection rates.}
   
   \authorrunning{Heywood et al.}
   \titlerunning{MESMER}
   
   \maketitle
   
%
%

\begin{figure*}
\vspace{200pt}
\includegraphics{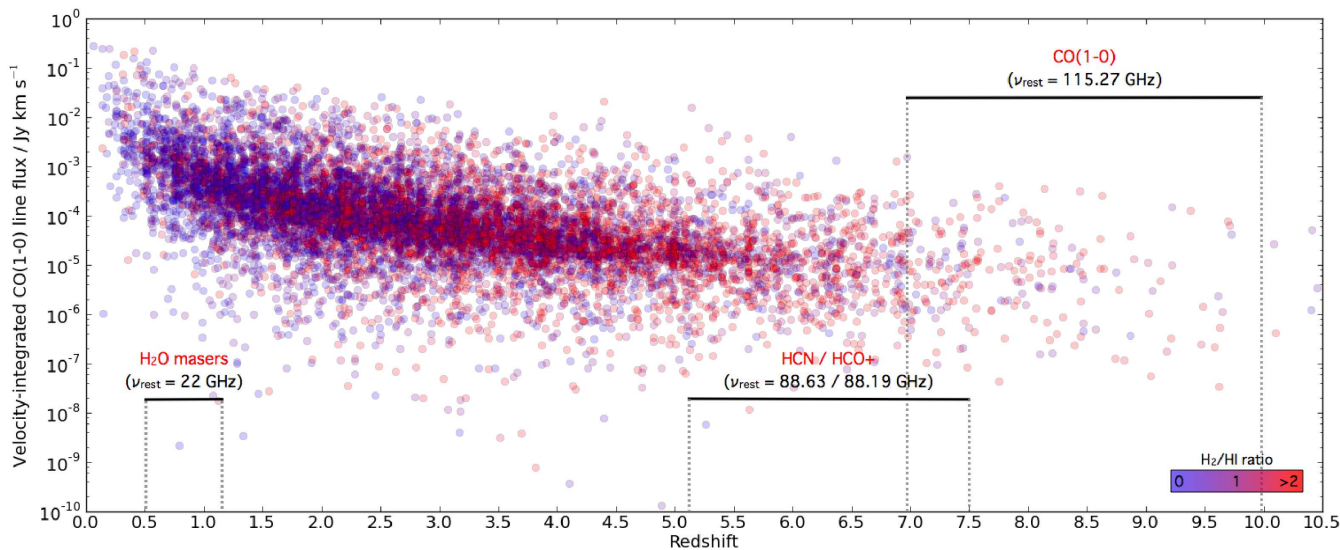}
\caption{The velocity-integrated CO($J$=1$\rightarrow$0) line flux for $\sim$10,000 simulated galaxies
   as a function of redshift. Key predictions of the S$^{3}$-SAX simulation are the 
   evolution with redshift of the line strength and the ratio of atomic to molecular gas (the latter represented by the colour gradient), and the cosmological decline in the
   source counts. The redshift ranges of some key spectral lines are plotted, as probed by an observation which
   covers 10.5 - 14.5 GHz. Note that the vertical positions of these markers has no significance.
        \label{fig:co_v_z}
}
\end{figure*}

\section{Introduction}

Carbon monoxide (CO) is the most abundant molecule after molecular hydrogen (H$_{2}$), however
the prohibitively high excitation temperature of the latter means that it is generally unobservable except in hot, shocked environments
(e.g. Rigopoulou, 2002). As such the CO molecule is generally used as a proxy tracer for H$_{2}$, with a CO/H$_{2}$ ratio within the Milky Way
of $\sim$10$^{-4}$. Observations of the CO molecule are thus critical in order to determine the cosmic evolution of the molecular gas 
density of the Universe ($\Omega_{H_{2}}(z)$). Since the molecular gas in a galaxy is the fuel reservoir for star-formation 
such observations are vital tools for studies of galaxy formation and evolution, and for 
tracing the star-formation history of the Universe. See the article by Walter et al. (2011) for a recent review 
of the importance of detecting molecular gas at high-z.

The usual targets for such observations are the rotational-vibrational transitions of the $^{12}$CO molecule. The $J$=1$\rightarrow$0 ground-state transition has a rest-frequency
of 115.27 GHz and higher-$J$ transitions occupy integer multiples of this frequency. As such, high-redshift observations are generally performed with millimetre-wave
interferometers, targeting $J$~$\rightarrow$~($J$-1) for $J$$\geq$1 (e.g. Tacconi et al., 2008), although the advent of the Expanded Very Large Array (EVLA) with its new Ka-band receivers (26 - 40 GHz)
has resulted in numerous recent detections of the ground-state transition of $^{12}$CO at high-z (e.g.~at z$\sim$2, Ivison et al., 2010; z$\sim$3, Riechers et al., 2010). 
Observations of the ground-state transition provide the best estimate of the \emph{total} amount of molecular gas within a galaxy. 
Typical targets for high-z observations to date have either been objects which are undergoing episodes of vigorous star-formation such as ``sub-mm (selected)
galaxies" (e.g.~Blain et al., 2002) with star-formation rates in excess of 1000~M$_{\odot}$~yr$^{-1}$, or luminous quasar host galaxies (e.g.~Riechers et al., 2008). 

It is desirable to observe multiple CO transitions in a given object in order to accurately determine the gas excitation state. Coupling this with the
need to perform such observations over a range of redshifts across cosmic time implies an observing frequency range which essentially spans two orders of magnitude,
from local, high-$J$ observations at close to 1~THz, to ground-state observations at 10~GHz which target objects at z$\sim$10, well within the cosmic
Epoch of Reionization (EoR; e.g.~Dunkley et al., 2009).

The MESMER\footnote{MeerKAT Search for Molecules in the Epoch of Reionization}
survey is one of the approved Large Survey projects for the forthcoming MeerKAT radio telescope, granted 6,500 hours of observing time, with the primary
goal of detecting large numbers of CO($J$=1$\rightarrow$0) emitters during the EoR. This article presents an overview of this observing programme, and couples
an existing simulation of the evolving molecular line properties of high-redshift galaxies with some instrumental considerations in order to
demonstrate the feasibility and predict the return of such observations.

\section{MeerKAT}

The MeerKAT\footnote{{\tt http://www.ska.ac.za}} array is currently under development and construction in the 
Karoo region of South Africa's Northern Cape province. It is one of the 
two complementary dish-based pathfinder instruments under construction by the 
two potential host nations of the core of the Square Kilometre Array (SKA), the other being 
the Australian SKA Pathfinder (ASKAP).

The current specification of MeerKAT describes 64 parabolic antennas 
with a diameter of 13.5 metres, and offset Gregorian optics. The 
unblocked aperture offered by such a system results in improved primary 
beam performance, and this optical configuration is a recent shift in 
design from having receivers at the prime focus, as is the case for 
KAT-7, the currently operational seven-element MeerKAT prototype. 
When complete, MeerKAT will be the most sensitive centimetre-wave instrument 
in the Southern Hemisphere, comparable in sensitivity to the EVLA, and much
faster in mapping speed by virtue of its smaller dishes and hence larger
field-of-view.

Final observing frequency specifications are likely to consist of three 
receiver bands: 0.58 -- 1.015 GHz, 1 -- 1.75 GHz and 8 -- 14.5 GHz, with a correlator capable of 
delivering complete coverage of the two lower bands and up to 4 GHz of instantaneous bandwidth for the high-band,
split over up to 25,200 frequency channels.

The proposed dish layout results in a fairly compact core, with over 50\% of 
the dishes within a 500 metre diameter and 75\% within a kilometre, and 
a maximum baseline of approximately 8 kilometres. A spur of antennas 
running along the access road (in an approximately easterly direction from the core) is also under consideration which would 
boost the maximum baseline to $\sim$50 kilometres.

\section{The S$^{3}$-SAX simulation}

The S$^{3}$-SAX simulation (Obreschkow et al., 2009a and references therein) 
is built upon the evolving dark matter skeleton of the Millennium Simulation
(Springel et al., 2005), and the galaxy semi-analytics of De Lucia \& Blaizot (2007), 
whereby the evolution of model galaxies placed at the centre of the dark matter haloes
was tracked. A physical recipe was applied to these two existing layers in order to
divide the cold gas in the model galaxies into H, H$_{2}$ and He, and realistic gas distributions
and velocity profiles were also added to the H and H$_{2}$.

Obreschkow et al. (2009b) subsequently applied a model to predict the luminosities of the first ten $J$ transitions 
of CO, including molecular cloud geometry and overlap, and contributions to gas heating from active galactic nuclei and starburst
activity, as well as the cosmic microwave background which becomes increasingly important at higher redshifts.

The simulation results in $\sim$30 million galaxies at z~=~0, each of which has a well-defined merger history traceable
to very high redshift, and a mass limit approximately equal to that of the Small Magellanic Cloud. The database
is publicly-accessible\footnote{{\tt http://s-cubed.physics.ox.ac.uk}} and contains amongst other things
the cosmologically evolving atomic and molecular line properties of every simulated galaxy.

Analytic expressions can be used to derive model line profiles from parameters within the database, and adding appropriate
instrumental noise to the line profile can allow us to make some predictions about the detectability of the line. This approach
is employed in order to obtain a first-order prediction for the return of the MESMER survey.

\section{MESMER}

The primary objective of the MESMER survey is to employ the high-frequency (14.5 GHz and below) receivers of 
MeerKAT to perform a systematic search for emitters of the $J$=1$\rightarrow$0 line of $^{12}$CO ($\nu_{rest}$~=~115.27 GHz).

   \begin{figure}
   \centering
   \includegraphics[width=0.9 \columnwidth]{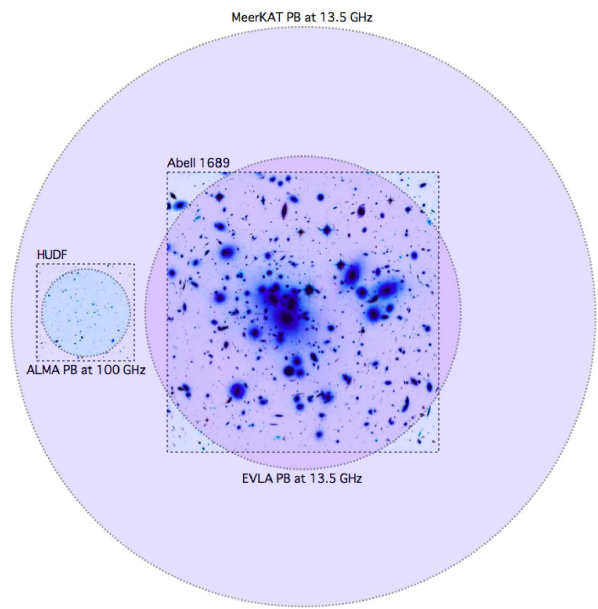}
 \caption{Approximate fields-of-view (full-width at half-maximum) for the ALMA, EVLA and MeerKAT interferometers, tuned to
   frequencies which target the CO molecule at z$\sim$7. The advantage of using cm-wave instruments for efficient line
   searches at this redshift should be immediately apparent. Also plotted within the squares for reference are the Hubble Ultra-Deep Field (Beckwith, 2006) and the Hubble
   Space Telescope observations of Abell 1689 (Benitez, 2002)}
            \label{fig:pb}
   \end{figure}

The upper limit of the high-frequency band at 14.5 GHz corresponds to CO($J$=1$\rightarrow$0) at z~=~6.95 which 
is inside the tail-end of the Epoch of Reionization. Using the wide bandwidth of MeerKAT to extend the observations
to lower frequencies pushes the targets further into the 
EoR. An observation with 4 GHz of bandwidth between 10.5 and 14.5 GHz 
would target CO($J$=1$\rightarrow$0) in the range 6.95 $\leq$ z $\leq$ 9.98, as delineated on Figure \ref{fig:co_v_z}.

The efficiency of a given survey is defined by the volume of space that can be covered and the flux limit that can be reached 
during the allotted time. Figure \ref{fig:pb} demonstrates the approximate size of the primary beam, full-width at half-maximum, for
three interferometers targeting CO at z$\sim$7. Observing at the lowest feasible frequency maximises the field of view, and to target CO at z$\sim$7, ALMA 
would be sensitive to the $J$=7$\rightarrow$6 transition at $\sim$100~GHz. Figure \ref{fig:pb} demonstrates the advantages that MeerKAT
has for covering large sky areas for observations of this nature. Also shown for reference are the Hubble Ultra-Deep Field (Beckwith, 2006) and the Hubble Space
Telescope observations of Abell 1689 (Benitez, 2002).

As is common practice for many survey programmes, a tiered or ``wedding cake" approach is adopted, whereby there
are a few distinct observing strategies, and a trade-off is made between sky coverage and survey depth between them. The three tiers
are described in Section \ref{sec:tiers}.

\section{Observing tiers and predicted detections}
\label{sec:tiers}

This section describes the current plan for the observing strategy, and how the specifications of MeerKAT can be combined with
the S$^{3}$-SAX simulation in order to predict the yield of a given observation.

Each pointing (or set of pointings) probes a volume of space defined by the field-of-view of the instrument (or size of the mosaic)
in the spatial directions, and by the bandwidth of the observation in the redshift direction. A sub-catalogue of galaxies within an equivalent volume
of space defined by a particular observation is then extracted from the S$^{3}$-SAX database by submitting a SQL query via
the web interface.

The root-mean-square (RMS) noise in an image produced from a single channel, with naturally-weighted visibilities, can be calculated
via the standard sensitivity equation (e.g.~Equation 6.62, Thompson, Moran \& Swenson, 2001):
\begin{equation}
\sigma = \frac{2  k_{B} T_{sys}}{A \eta_{Q} n_{p} \sqrt{N (N-1) \Delta \nu \tau }}
\end{equation}
where $k_{B}$ is the Boltzmann constant, $T_{sys}$ is the system temperature (50~K), $A$ is the effective area of the 13.5-metre diameter antennas, 
$\eta_{Q}$ is an efficiency term due to correlator quantization losses (0.95), $n_{p}$ is the number of averaged orthogonal
polarization products (2), $N$ is the number of antennas
(64) $\Delta \nu$ is the channel width (3.8~MHz, i.e. 1024 channels for 4 GHz of total bandwidth) and $\tau$ is the on-source time.

The analytic line profiles for each galaxy in the sub-catalogue are then evaluated sequentially. A source is considered to be detected if \emph{(a)}
the line flux integrated over the channel width at the peak of the line exceeds 5$\sigma$; and \emph{(b)} the full-width of the line at the
half-maximum level exceeds three channel widths. The threshold value of 5$\sigma$ suggests that the resulting source counts will be conservative,
however this may be offset by the assumption that the data can be perfectly calibrated.

\subsection{Tier 3: a square-degree survey}

The broadest and shallowest observing tier is likely to be a square-degree survey over
a field with excellent multiwavelength coverage (e.g.~the Extended Chandra Deep-Field South).
It is anticipated that 2,000 hours will be dedicated to this. With a traditional hexagonal
mosaic pattern and an assumed calibration overhead of 20\%, this results in approximately 340
pointings with 4.7 hours of actual on-source time each.

The channel RMS for such an observation is 32~$\mu$Jy, and the extracted square-degree of the S$^{3}$-SAX
simulation at 6.95 $\leq$ z $\leq$ 9.98 (4~GHz bandwidth) contains 87,597 galaxies. Using the method outlined above
suggests that such an observation would detect 186 CO($J$=1$\rightarrow$0) emitters. Note that no primary beam effects are taken into account 
as the mosaic pattern should allow coverage of the field
with approximately uniform sensitivity.

\begin{figure}
   \centering
 \includegraphics[width =\columnwidth]{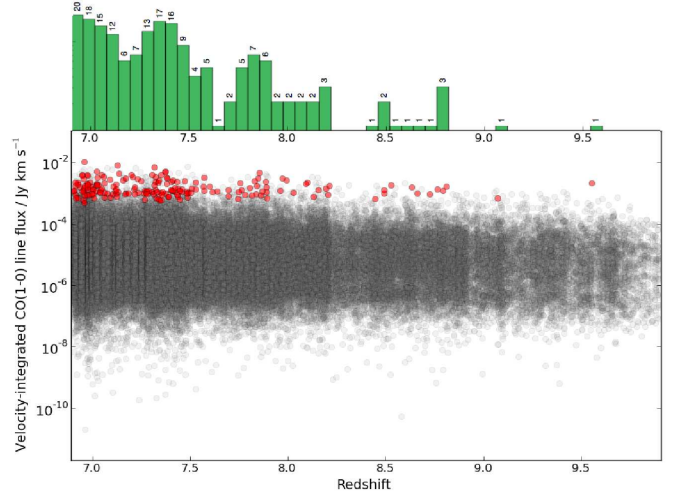} 
 \caption{The velocity integrated CO($J$=1$\rightarrow$0) line fluxes plotted as a function of redshift from the simulated galaxies
   within a volume of space defined by one square degree and 6.95 $\leq$ z $\leq$ 9.98. Detections are determined by the
   criteria outlined in Section \ref{sec:tiers} and these are marked in red. A histogram (with a logarithmic y-axis) of the sources in 50 redshift bins
   is plotted at the top of the figure.}
            \label{fig:sq_deg}
   \end{figure}

Figure \ref{fig:sq_deg} shows the velocity-integrated CO($J$=1$\rightarrow$0) line fluxes for the simulated sources as a function of redshift. 
Detections are marked in red, and binning these detections by redshift forms the green histogram at the top of the figure. 

\subsection{Tier 2: quasars and Lyman-$\alpha$ emitters}

Tier 2 will consist of 9 pointings, each having a total observing time of 300 hours, which with the calibration overhead
results in 240 hours of on-source time. The channel RMS in this case is
5.3~$\mu$Jy. Targets are likely to be the highest-redshift quasars, regions containing
strong overdensities of Lyman-$\alpha$ emission, and regions selected for future, deep infrared observations with the James Webb Space Telescope.
Although the current highest-redshift quasars are outside of the MeerKAT frequency range, 
Willott et al. (2010) using current constraints on the quasar luminosity function at z$>$6 predict that
the VISTA survey programmes will detect $>$20 quasars at z$>$6.5, and $>$25\% of these are 
expected to fall within the redshift range observable by MeerKAT.

The most luminous quasars are likely to correspond to the most molecule-rich galaxies in the early Universe (Walter et al., 2009)
and narrowband search techniques which discover overdense regions of Lyman-$\alpha$ emission are likely to be revealing the sites
of primordial cluster formation (Ouchi, 2005). 

It is likely that such deep observations will not only reveal the molecular gas of the primary target, but will also contain many other CO-line
emitters. Selecting several random volumes of space corresponding to a single pointing with 4~GHz of bandwidth suggests that each pointing is likely
to yield a total of $\sim$20 such detections.

The primary beam of an interferometer causes a drop in sensitivity away from the pointing centre and this is approximately accounted for when simulating this tier.
A normalised Gaussian with a FWHM corresponding to the size of the Airy disk for a 13.5~m aperture is centred on the pointing direction. This then
defines the function by which source line fluxes are attenuated based on their position in the field.

\subsection{Tier 1: strong gravitational lenses}

The deepest pointings will target three strong gravitational lenses. Potential targets are Abell 370, Abell 1689 and MACS2129
which have no bright radio galaxies in the field. The magnification afforded by the foreground cluster will amplify the
strength and counts of the background high-redshift population.
Each pointing will have 600 hours of observing time (480 hours on-source), giving a
channel RMS of 3.8~$\mu$Jy.  

Mass models for the above galaxy clusters applied to the dN/dz counts extracted from the S$^{3}$-SAX simulation suggest an
average magnification factor of $\sim$2 across the field, with high factors ($\>$5--30) in small areas of the source plane.
It is estimated that the overall detection rate for high-z CO emitters for these observations will be $\sim$30--40 galaxies
per pointing. 

In the small regions of very high magnification galaxies with integrated line fluxes of $\sim$2$\times$10$^{-5}$~Jy~km~s$^{-1}$ corresponding
to objects with 
H$_{2}$ masses as low as $\sim$10$^{8}$~M$_{\odot}$ and very modest levels of star formation ($\sim$5~M$_{\odot}$~yr$^{-1}$)
start to become detectable. The boosting of fluxes at the higher end of the luminosity function emphasises the importance
of targeting several lens fields to account for cosmic variance.

Note also (with reference to Figure \ref{fig:pb}) that the field of view of MeerKAT is sufficient to cover the whole strongly-lensed area of such clusters
in one shot. There is no need to divide the observing time over multiple pointings with painstaking targeting of the lens caustics.

\section{Parallel science goals}

Very deep, high-frequency observations as will be generated by this programme will obviously have a wealth of other applications.
In addition to the primary goal of detecting molecular line emission during the EoR there will also be the possibility to search for other
spectral lines at lower redshifts. 

Hydrogen cyanide (HCN($J$=1$\rightarrow$0), $\nu_{rest}$~=~88.63 GHz) and formylium (HCO$^{+}$($J$=1$\rightarrow$0), 
$\nu_{rest}$~=~89.19 GHz) lines will be 
visible for the redshift range 5.1~$\leq$~z~$\leq$~7.5, corresponding to 14.5--10.5 GHz.  
These lines are tracers of dense (n$_{H_{2}}$ $>$ 10$^{(11-12)}$~m$^{-3}$)
regions of gas, which are typically associated with the star-forming cores
of giant molecular clouds (see e.g. Solomon and Vanden Bout, 2005), and the ratio of these
lines is often used as an indicator of starburst versus AGN activity in galaxies (e.g. Imanishi
et al., 2007; Knudsen et al., 2007).

Despite these lines being significantly weaker than CO by a factor of 10--40, 
predictions from Carilli \& Blain (2002) show that a small 
fraction of the detected high-z CO($J$=1$\rightarrow$0) lines will be false detections corresponding to HCN at lower-z.  
Those authors estimate a detection rate for HCN (expressed as a percentage of CO 
detections) of 1.5\% at z = 1, rising to 5\% at z = 3.

Separating these line sources from the genuine z$>$7 CO detections can be accomplished by identifying
both HCN and HCO$^{+}$ in the same spectrum due to their proximity in frequency. For detections
where only a single line is visible then near-infrared identifications and photometric redshifts are likely to be necessary. Failing that,
follow-up observations (e.g. with ALMA) will confirm the redshift by targeting other spectral lines.

Additionally the data will facilitate a deep search for 0.5~$\leq$~z~$\leq$~1.2 water masers ($\nu_{rest}$~=~22~GHz). 
Currently, the two most distant detections are at z$\sim$2.5 and z$\sim$0.66 (Impellizzeri et al. 2008; 
Barvainis \& Antonucci 2005) the 
former being gravitationally lensed by a factor $\mu$~$\simeq$~33. The redshift range probed by MESMER 
provides an ambitious yet practical next step to push the H$_{2}$O luminosity function out to higher 
z. Any detections will be followed up with VLBI to measure robust kinematic black hole masses 
and independent geometric distances, provided the maser emission is sufficiently luminous when 
resolved. Independent distance measurement of course provide an estimate of the Hubble parameter 
in a cosmologically significant redshift range where dark energy becomes comparable to the matter 
energy-density in the Universe.

The redshift ranges for the two parallel spectral line surveys described above are marked on Figure \ref{fig:co_v_z}.

The MESMER programme will also yield very deep $\sim$12~GHz continuum data. Spectral indices determined by combining these data with
those of lower-frequency radio observations (e.g.~from the MIGHTEE Large Survey project, PIs van der Heyden \& Jarvis) will be very useful for separating the synchrotron from the free-free emission in star-forming galaxies,
and for probing AGN / starburst composites.

\section{Imaging simulation pipeline}

The next step in terms of simulating the observations described in this document is to move from the simple analytic approach to a full imaging
simulation, and a software pipeline has been assembled for this purpose. The pipeline takes as its input a subcatalogue of galaxies from S$^{3}$-SAX
and a list of observational parameters, detailing such things as the observation duration, the pointing direction and the frequency setup. 
The basic steps in the simulation chain are as follows: \emph{(a)} generation of an ``ideal'' radio sky from the simulation database using the {\tt S3Tools} (Levrier et al., 2009); \emph{(b)}
pixelwise attenuation of the sky model by an analytic primary beam model; \emph{(c)} use of the {\tt CASA}\footnote{{\tt http://casa.nrao.edu}} {\tt sm} tool to generate
a per-channel Measurement Set, to serve as an empty ``crate'' for visibilities; \emph{(d)} use of the {\tt MeqTrees} package (Noordam \& Smirnov, 2010) to invert the
sky model and fill the Measurement Sets with appropriately-noisy visibilities; \emph{(e)} imaging and (optional) deconvolution of the visibilities using the {\tt lwimager},
part of the {\tt casarest} package; and \emph{(f)} (optional) correction of the primary beam model. 

This process can operate from start to finish without user intervention using a combination of shell and Python scripts, and by operating {\tt MeqTrees} in batch mode. 
Each frequency channel is treated as an individual simulation, which naturally lends itself to the jobs being split over multiple processors. The final set of channel maps can also 
be optionally combined into a single FITS cube.

\begin{figure}
   \centering
\includegraphics[width=\columnwidth]{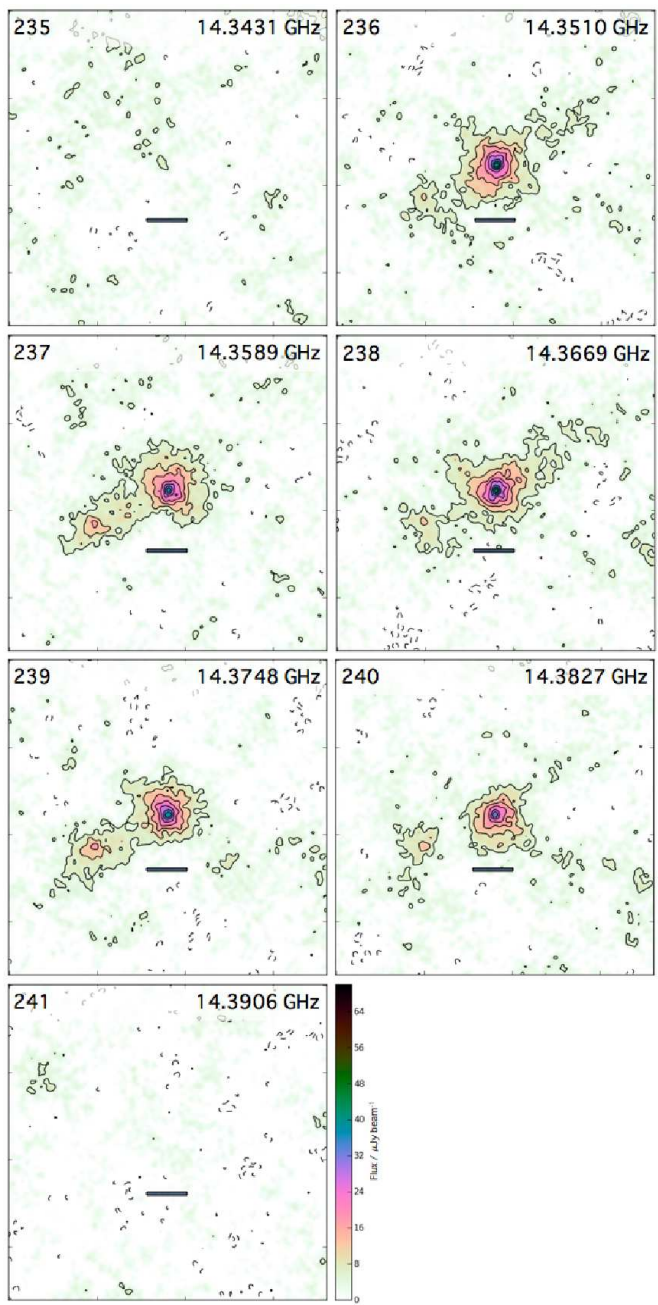}
\caption{A 180 $\times$ 180 $\times$ 7 subcube extracted from the full size single-pointing simulation, showing the detection of CO($J$=1$\rightarrow$0) emission from a pair of z~=~7.02
   simulated galaxies. Channel number and frequency are marked on each frame. The horizontal bar on each frame spans 5 arcseconds.}
            \label{fig:mesmer_detection}
   \end{figure}

This software chain has been used to simulate a single 480-hour pointing with 2 GHz of bandwidth (12.5 -- 14.5 GHz), consistent with a MESMER tier 1 observation of high-z CO($J$=1$\rightarrow$0), with 2:1 channel averaging. Note however that the effects of the foreground lensing
clusters are not (yet) included. The end result is a 4096 $\times$ 4096 $\times$ 256 pixel datacube. The final FITS cube is 17 GB although the whole process generates approximately
0.3 TB of data, mostly visibilities. A solid detection from within this simulated cube, showing the CO($J$=1$\rightarrow$0) emission from a pair of galaxies at z~=~7.02, is presented in Figure \ref{fig:mesmer_detection}.

The goal of this work, in addition to obtaining a more robust estimate of the detection rate, is to test the efficacy of source finder software and spectral-line stacking code 
in the presence of known noise and sky properties for MeerKAT-scale datasets. 
A separate paper by Heywood et al. (2011; in prep) will describe this work in full.

Note that although the simulation method is described here in the context of MeerKAT it can be used to simulate completely arbitrary interferometers, 
and {\tt MeqTrees} allows for the introduction of complex instrumental effects (e.g.~more accurate primary beam treatment from electromagnetic dish simulations) 
without having to develop code beyond the Python layer. 
Additionally, the sky model can also be completely arbitrary, 
but without modification to this collection of scripts, it  can be generated from the HI component of S$^{3}$-SAX, or any of its other CO transitions, and also the S$^{3}$-SEX continuum simulation\footnote{Also 
available at {\tt http://s-cubed.physics.ox.ac.uk}} devised by Wilman et al. (2008).
The article by Levrier et al. (2009) describes in more detail the possibilities for generating sky models from the S$^{3}$ suite of simulations.

\section{Conclusions}

The MESMER Large Survey project will utilise the high-frequency receivers of the forthcoming MeerKAT array and, using a 
combination of very deep single pointings and shallower wide-area observations totalling 6,500 hours, is expected to detect the $J$=1$\rightarrow$0 line of the $^{12}$CO
molecule in $\sim$400 galaxies in the range 6.95~$\leq$~z~$\leq$~9.98, during the cosmic Epoch of Reionization.

The survey will naturally include parallel spectral line surveys at other redshifts, notably a HCN / HCO$^{+}$ at
5.1~$\leq$~z~$\leq$~7.5, and H$_{2}$O masers at 0.5~$\leq$~z~$\leq$~1.2, as well as deep $\sim$12~GHz continuum maps
which will be valuable for AGN / starburst studies.

Making predictions about the feasibility of such observations (and thus getting a handle on the science return) in
addition to investigating via simulation the instrumental performance of current and next-generation 
interferometers was the primary motivation for the development of the S$^{3}$-SAX simulation (and the other S$^{3}$ simulations).

This article has presented both a simple, first-order application of the simulation via the generation of analytic line profiles, and also a framework
for extending the instrumental simulations into a full interferometric imaging simulation. The latter approach is particularly powerful
for assessing current instrumental designs, testing software on large-scale datasets, and naturally provides a more robust prediction
as to the high-z molecular line detections that can be made with the MeerKAT array.

The framework used for this investigation can also be used to simulate other key science, including extragalactic neutral hydrogen
and continuum.

\begin{acknowledgements}
The authors with to thank Bradley Frank for providing the most recent proposed MeerKAT antenna positions, and 
Dave Green for devising and sharing the ``cube helix" colourmap, an inversion of which was used to
produce Figure \ref{fig:mesmer_detection}.
\end{acknowledgements}

\end{document}